\documentclass[twocolumn,showpacs,prl]{revtex4}
\usepackage{graphicx,epsfig}

\begin{document}

\noindent
DESY 03-172, TTP03-40, UB-ECM-PF-03-28\hfill 

\title{
\boldmath 
$M(\eta_b)$  and $\alpha_s$ from Nonrelativistic Renormalization Group
\unboldmath}
\author{Bernd A. Kniehl,$\!{}^1$ Alexander A. Penin,$\!{}^{2,3}$
Antonio Pineda,$\!{}^4$
Vladimir A. Smirnov,$\!{}^{5,1}$ Matthias Steinhauser$^1$}
\affiliation{
$^1$ II. Institut f\"ur Theoretische Physik, Universit\"at Hamburg,
22761 Hamburg, Germany\\
$^2$ Institut f{\"u}r Theoretische Teilchenphysik,
Universit{\"a}t Karlsruhe, 76128 Karlsruhe, Germany\\
$^3$ Institute for Nuclear Research, Russian Academy of Sciences,
117312 Moscow, Russia\\
$^4$ Dept. d'Estructura i Constituents de la Mat\`eria, U. Barcelona,  
E-08028 Barcelona, 
Spain\\
$^5$ Institute for Nuclear Physics, Moscow State University, 119899 Moscow,
Russia}

\date{\today}

\begin{abstract}
We sum up the next-to-leading logarithmic corrections to the 
heavy-quarkonium hyperfine splitting using the nonrelativistic
renormalization group. On the basis of this result, we predict the
mass of the $\eta_b$ meson to be 
$M(\eta_b)=9419 \pm 11\,{(\rm th)} \,{}^{+9}_{-8}\,(\delta\alpha_s)~{\rm MeV}$.
The experimental measurement of  $M(\eta_b)$ with a few MeV error
would be sufficient to  determine  $\alpha_s(M_Z)$ with 
an accuracy of $\pm 0.003$. 
The use of the nonrelativistic renormalization  group  is 
mandatory to  reproduce the experimental value  of the hyperfine splitting 
in charmonium.
\end{abstract}

\maketitle

The theoretical study of nonrelativistic heavy-quark-antiquark systems is
among the earliest applications of perturbative quantum chromodynamics (QCD)
\cite{AppPol} and has by now become a classical problem. 
Its applications to bottomonium or toponium
physics entirely rely on the first principles of QCD.
This makes heavy-quark-antiquark systems an ideal laboratory to determine
fundamental parameters of QCD, such as the strong-coupling constant 
$\alpha_s$ and the heavy-quark masses $m_q$.
Besides its phenomenological importance, the heavy-quarkonium system is also
very interesting from the theoretical point of view because it possesses a
highly sophisticated multiscale dynamics and its study demands the full power
of the effective-field-theory approach.
The properties of the $\Upsilon$ mesons, the  bottom quark-antiquark 
spin-one bound states,  are measured experimentally with great precision, and
recent theoretical analysis of the $\Upsilon$ family based on  high-order 
perturbative calculations  resulted in determinations of the bottom-quark mass 
$m_b$ with unprecedent accuracy \cite{KPP,PinYnd,PenSte}. 

In contrast to the $\Upsilon$ family,
the current experimental situation with the spin-zero $\eta_b$ meson 
is rather uncertain: only one candidate event in 
$\gamma\gamma\to\eta_b$ production has been detected so far, 
which, however, is consistent with the expected background~\cite{Hei}. 
Yet, the discovery of  the $\eta_b$ meson is one of the primary goals of the CLEO-c 
research program \cite{Sto}.
An accurate prediction of its mass $M(\eta_b)$ is thus a big challenge
and a test for the QCD theory of heavy quarkonium. 
Moreover, the hyperfine splitting (HFS) of the  bottomonium ground 
state,  $E_{\rm hfs}=M(\Upsilon(1S))-M(\eta_b)$,
is  very  sensitive to   $\alpha_s$ and, 
with the advancement of the experimental measurements, 
could become a  competitive source for 
the determination of the strong coupling constant.  

The HFS in quarkonium has been a subject of several theoretical 
researches \cite{BucNg}.  
To our knowledge, the  next-to-leading order (NLO)
${\cal O}(\alpha_s)$ correction is currently known in a closed 
analytical form only for the ground state HFS \cite{PenSte}.
In this letter, we  generalize this result to the excited states
and present the  analytical renormalization-group-improved
expression for  the heavy-quarkonium HFS  in the next-to-leading logarithmic 
(NLL) approximation,
which sums up all the corrections of the form $\alpha_s^n\ln^{n-1}\alpha_s$.
We apply it to predict $M(\eta_b)$.
The result can  be used for extracting  $\alpha_s(M_Z)$ 
from  future experimental data on the $\eta_b$ meson mass.
  
The leading-order (LO) result for the HFS is proportional to the fourth power
of $\alpha_s$, $E^{LO}_{\rm hfs}=
C_F^4\alpha_s^4(\mu)m_q/(3n^3)$, where $C_F=(N_c^2-1)/(2N_c)$,
and suffers from a strong dependence on the renormalization
scale $\mu$ of $\alpha_s(\mu)$, which essentially limits the numerical
accuracy of the approximation.
Thus, the proper fixing of  $\mu$ is mandatory
for the HFS phenomenology. 
The  scale dependence of a finite-order result is canceled against 
the higher-order
logarithmic contributions proportional to a power of $\ln(\mu/\bar\mu)$,
where $\bar\mu$ is a dynamical scale of the nonrelativistic
bound-state problem. 
The {\it physical} choice of the scale  $\mu=\bar\mu$  eliminates these 
potentially  large logarithmic terms and {\it a priori} minimizes
the scale dependence. 
However, the dynamics of the  nonrelativistic  bound state  is characterized
by three well separated scales: the hard scale of the heavy-quark mass
$m_q$, the soft scale of the bound-state momentum  $vm_q$, 
and the ultrasoft scale  of the bound-state energy $v^2m_q$,
where $v\propto\alpha_s$ is
the velocity of the heavy quark inside the approximately Coulombic
bound state.  To make the procedure of scale fixing self-consistent,
one has to resum to all orders the large logarithms of the scale
ratios characteristic for the nonrelativistic bound-state problem.  
The  resummation of the logarithmic corrections requires an appropriate
conceptual framework.  The effective field theory \cite{CasLep} is now
recognized as a powerful tool for the analysis of  multiscale
systems, which is at the heart of the recent progress in the
perturbative QCD bound-state calculations. The main idea of this method
is to decompose the complicated multiscale problem into a sequence of
simpler problems, each involving a smaller number of scales.  The
logarithmic corrections originate from  logarithmic integrals over
virtual momenta ranging between the scales and reveal themselves as the
singularities of the effective-theory couplings. The renormalization
of these singularities allows one to derive the equations of the {\it
nonrelativistic renormalization group} (NRG), which describe the
running of the effective-theory couplings, {\it i.e.} their dependence
on the effective-theory cutoffs.  The solution of these equations sums up
the logarithms of the scale ratios.

To derive the NRG equations necessary for the NLL analysis of the HFS, we rely
on the method based on the formulation of the nonrelativistic
effective theory known as potential NRQCD (pNRQCD) \cite{PinSot1}. The
method was developed in Ref.~\cite{Pin} where, in particular, the leading
logarithmic (LL) result for the HFS has been obtained (see also Ref.~\cite{HMS}).  
A characteristic feature of the NRG is the correlation of the dynamical 
scales, which leads to the correlation of the cutoffs \cite{LMR}.
For  perturbative calculations within the 
effective theory, dimensional regularization is used to
handle the divergences, and the formal expressions derived from the
Feynman rules of the effective theory are understood in the sense of
the threshold expansion \cite{BenSmi}.  This approach
\cite{KPSS,PinSot2,CMY,KniPen1} possesses two crucial virtues: the absence of
additional regulator scales and the automatic matching of the
contributions from different scales.

Let us give a few details of the NLL analysis. 
We distinguish the
soft, potential, and ultrasoft anomalous dimensions corresponding to
the ultraviolet divergences of the soft, potential, and ultrasoft {\it
regions} \cite{BenSmi}.  The LL approximation is determined by the
one-loop soft running of the effective Fermi coupling $c_F$ and the
spin-flip four-quark operator \cite{Pin}. 
In the NLL approximation, all three types of running contribute. 
We need  the two-loop soft running of  $c_F$, which is known \cite{ABN},
and the two-loop soft running  of the  spin-flip
four-quark operator, which we compute by adopting the
technique used in Ref.~\cite{KPSS} for the calculation of the 
two-loop $1/(m_qr^2)$ non-Abelian potential. 
To compute the potential running,
we inspect all operators that lead to spin-dependent
ultraviolet divergences in the time-independent perturbation theory
contribution with  one and two potential loops \cite{KniPen2,Pin}.
They include (i)  the ${\cal
O}(v^2,\alpha_sv)$ operators \cite{KPP,PinYnd}, (ii) the tree ${\cal
O}(v^4)$ operators, some of which can be checked against the QED analysis
\cite{CMY}, and (iii) the one-loop ${\cal O}(\alpha_sv^3)$ operators, for which
only the Abelian parts are known \cite{CMY}, while the non-Abelian parts
are new.  In the NLL approximation, we need the LL soft and ultrasoft
running of the  ${\cal O}(v^2)$ and  ${\cal O}(v^4)$
operators, which enter the two-loop time-independent
perturbation theory diagrams,
and the NLL  soft and ultrasoft running of  the ${\cal
O}(\alpha_sv)$  and ${\cal O}(\alpha_sv^3)$ operators,
which contribute at one loop. 
The running of the ${\cal O}(v^2,\alpha_sv)$  operators is already known
within pNRQCD \cite{Pin}. The running of the other operators is new. For
some of them, it can be obtained using  reparameterization invariance 
\cite{Manohar}.

Besides the running discussed above, we need the initial
conditions for the NRG evolution given by the known one-loop result
\cite{BucNg}.  With the anomalous dimensions and initial conditions at
hand, it is straightforward to solve the system of the nonlinear
differential equations for the effective couplings and get the NLL
result for the HFS.  The corresponding expression for general color (light
flavor) number $N_c$ ($n_l$) and for arbitrary principal quantum
number $n$ is too lengthy to be shown in this Letter, so we
present the explicit analytical expression only for $N_c=3$, $n_l=4$,
and $n=1$, which applies to the bottomonium ground state. It reads
\begin{eqnarray}
\nonumber
\lefteqn{
E_{\rm hfs}^{NLL} = {C_F^4\alpha_s^4(\mu)m_b \over 3}
\left\{
\frac{27}{14}y^{-1}-\frac{13}{14}y^{-\frac{18}{25}}
\right.}
\\
\nonumber
&&
+{\alpha_s(m_b)\over\pi}\left[\left( 
\frac{1037}{224}  
+\frac{3938372260247\,\pi^2}{256171608000} 
-\frac{3}{4}\ln{2}\right)y^{-1}
\right.
\\
\nonumber
&&
-\frac{1024\,\pi^2}{143}\,y^{-\frac{39}{50}} - 
\left(  
\frac{102973}{26250} +
\frac{184336\,\pi^2}{25725}
\right)\,y^{-\frac{18}{25}}
\\
\nonumber
&&
+\frac{1024\,\pi^2}{675}y^{-\frac{1}{2}} 
+\frac{671\,\pi^2}{1029}y^{-\frac{11}{25}}
-\frac{3\,\pi^2}{23}y^{-\frac{2}{25}}
\\
\nonumber
&&
+\left(
-\frac{13427921}{1260000}
+ 
\frac{161939\,\pi^2}{302400}
\right)y^{\frac{7}{25}}
+\frac{4\,\pi^2}{41}y^{\frac{16}{25}}
\\
\nonumber
&&
+\frac{1377}{56} 
-\frac{233027\,\pi^2}{45500}  
+\frac{{\pi}^2}{75}y 
\\
\nonumber
&&
+\frac{6\,\pi^2}{25} y^{-1}
\left(B_{y/2}
\left(\frac{32}{25},\frac{43}{25}\right)-B_{1/2}
\left(\frac{32}{25},\frac{43}{25}\right)\right)
\\
\nonumber
&&
-\frac{2873\,\pi^2}{7182} 
y^{\frac{32}{25}}{{}_2F_1}
\left(\frac{57}{25},1;\frac{82}{25};\frac{y}{2}\right)
\\
\nonumber
&&
+\frac{2873\,\pi^2}{3591} 
y^{-1}{{}_2F_1}
\left(1,1;\frac{82}{25};-1\right)
+ \left(
\frac{675}{28} - 
\frac{533}{42}y^{\frac{7}{25}}
\right)
\\
\nonumber
&&
\times\ln \left(\frac{{\mu}}{{\bar \mu}}\right)
+\frac{85248\,\pi^2}{30625}y^{-1}\ln y
+\left(-\frac{42432}{4375}y^{-1} 
\right.
\\
&&
\left.
\left.
\left.
+\frac{21216}{4375}-\frac{2873}{1575}y^{\frac{7}{25}} 
\right)\pi^2\ln(2-y) 
\right]\right\}\,,
\label{resum}
\end{eqnarray}
where $\alpha_s$ is
renormalized in the $\overline{\rm MS}$ scheme,
$y=\alpha_s(\mu)/\alpha_s(m_b)$,
$\bar\mu=C_F\alpha_s(\mu)m_b$, $B_{z}(a,b)$ is the
incomplete beta-function, ${}_2F_1(a,b;c;z)$ is the
hypergeometric function, and
${}_2F_1(1,1;82/25;-1)=0.7875078\ldots$.  By expanding the
resummed expression up to ${\cal O}(\alpha_s^2)$, we get
\begin{eqnarray}
  \lefteqn{E_{\rm hfs}^{NLL} = E_{\rm hfs}^{LO}\bigg\{
  1 + \frac{\alpha_s}{\pi}\bigg[
    \bigg( 
    \frac{1}{2}
    +\frac{5}{4n}
    -n\Psi_2(n)
    \bigg)\beta_0}
    \nonumber\\&&\mbox{}
    +\bigg(
    -\frac{37}{72}
    -\frac{7}{8n}
    +\frac{7}{4}\left(\Psi_1(n+1)+\gamma_E + L_{\alpha_s}^n \right)
    \bigg) C_A
    \nonumber\\&&\mbox{}
    -\frac{C_F}{2}
    +\frac{3}{2}(1-\ln 2) T_F -\frac{5}{9}n_l T_F
    \bigg]
    \nonumber\\&&\mbox{}
  + \left(\frac{\alpha_s}{\pi}\right)^2 L_{\alpha_s}^n \bigg[
    \bigg(\frac{19}{6}C_A^2-\frac{5}{6}C_An_l T_F\bigg) 
    L_{\alpha_s}^n
    + \frac{\beta_1}{8}
    \nonumber\\&&\mbox{}
    + \bigg(
    \bigg(
    \frac{169}{144} + \frac{2}{n}
    +\frac{3}{8}\left(\Psi_1(n+1)+\gamma_E\right)
    -\frac{7}{4} n\Psi_2(n)
    \bigg) C_A
    \nonumber\\&&\mbox{}
    -C_F + \frac{3}{2}(1-\ln 2)T_F 
    \bigg)\beta_0
    + \bigg(
    -\frac{23}{27} 
    -\frac{\pi^2}{6}
    - \frac{7}{8n}
    \nonumber\\&&\mbox{}
    + \frac{7}{4}\left(
    \Psi_1(n+1)+\gamma_E
    \right)
    \bigg)C_A^2
    + \bigg(\frac{7}{4}-\frac{11\pi^2}{8}\bigg) C_A C_F
    \nonumber\\&&\mbox{}
    - \frac{\pi^2}{2}C_F^2
    - \frac{44}{27} C_AT_F n_l
    + \frac{1}{2}C_F T_F n_l     
    \bigg]
  \bigg\}
  \label{expan}
  \,,
\end{eqnarray}
where $\alpha_s\equiv \alpha_s(\mu)$, $\mu=\bar{\mu}/n$, $C_A=N_c$, $T_F=1/2$, 
$\beta_i$ is the $(i+1)$-loop coefficient of the QCD $\beta$ function
($\beta_0=11C_A/3-4T_Fn_l/3,\ldots$),
$L^n_{\alpha_s}=\ln\left(C_F\alpha_s/n\right)$,
$\Psi_n(x)=d^n\ln\Gamma(x)/dx^n$, $\Gamma(x)$ is Euler's $\Gamma$
function, and $\gamma_E=0.577216\ldots$ is Euler's constant. In
Eq.~(\ref{expan}), we keep the full dependence on $N_c$, $n_l$, and $n$.
The ${\cal O}(\alpha_s^2\ln^2\alpha_s)$ term is known \cite{Pin,HMS}, while the
${\cal O}(\alpha_s^2\ln\alpha_s)$  term is new. 

\begin{figure}
\epsfxsize=8.5cm 
\epsfig{figure=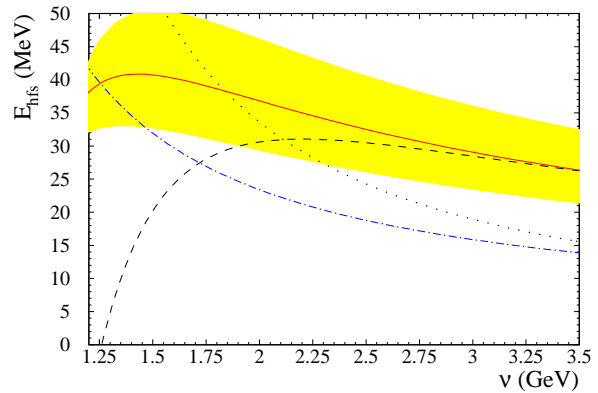,height=6cm}
\caption{\label{fig1} HFS of 1S bottomonium as a function of the
renormalization scale $\mu$ in the LO (dotted line), NLO (dashed line), LL
(dot-dashed line), and NLL (solid line) approximations. For the NLL
result, the band reflects the errors due to $\alpha_s(M_Z)=0.118\pm
0.003$.}
\end{figure}

For the numerical estimates, we adopt the following strategy.  First, we
compute the dimensionless ratio $E_{\rm hfs}/M(\Upsilon(1S))$, which only
has logarithmic dependence on $m_b$. At the order of interest,
we have $M\left(\Upsilon(1S)\right)=2m_b$. 
Then, the ratio is converted into $E_{\rm hfs}$ by multiplying it
with the physical $\Upsilon(1S)$ meson mass,
$M\left(\Upsilon(1S)\right)=9460.30\pm 0.25$~MeV \cite{Hag}.  In
Fig.~\ref{fig1}, the  HFS for the bottomonium ground state is plotted as a
function of $\mu$ in the LO, NLO, LL, and NLL
approximations.  As we  see,  the LL curve shows  a weaker 
scale dependence compared to the LO one. 
The scale dependence of the NLO and NLL expressions is further reduced,  
and, moreover, the NLL approximation remains stable at 
the physically motivated  scale of the inverse Bohr radius,
$C_F\alpha_s m_b/2\sim 1.5$~GeV, where the fixed-order expansion 
breaks down. At the scale
$\mu'\approx 1.3$~GeV, which is close to the inverse Bohr radius, the NLL
correction vanishes as one expects from  general arguments.
Furthermore, at $\mu''\approx 1.4$~GeV, the result becomes independent
of $\mu$; {\it i.e.}, the NLL curve shows a local maximum.  This
suggests a nice convergence of the logarithmic expansion
despite the presence of the ultrasoft contribution with $\alpha_s$
normalized at the rather low scale $\bar\mu^2/m_b\sim 0.8$~GeV.  By taking 
the difference of the NLL and LL results at the local maxima 
as a conservative estimate of the error due to uncalculated higher-order
contributions,  we get $E_{\rm hfs}=41\pm 8$~MeV. 

So far, we restricted the analysis to  purely perturbative
calculations.  The nonperturbative contribution to the HFS can in
principle be investigated by the method of vacuum condensate expansion
\cite{VolLeu}. The resulting series does not converge well and suffers
from large numerical uncertainties \cite{TitYnd2}.  A reliable
quantitative estimate of the nonperturbative contributions to the HFS can
be obtained by comparison with lattice simulations.  The HFS in
bottomonium was studied on the lattice by several groups.  The SESAM
Collaboration \cite{Eic} reported the value $E_{\rm hfs}=33.4\pm 1.9$~MeV,
while the CP-PACS Collaboration \cite{Man} obtained $E_{\rm hfs}=33.2\pm 1.0$~MeV
(only the  statistical errors are quoted).
Both groups used unquenched nonrelativistic lattice QCD.  The result
from quenched relativistic QCD on an anisotropic lattice \cite{Lia} is
$E_{\rm hfs}=33.8\pm 3.1$~MeV. The central values are in good agreement and
undershoot our NLL result by approximately 7~MeV, which we take as an
estimate of the nonperturbative contribution.

\begin{figure}
\epsfxsize=8.5cm 
\epsfig{figure=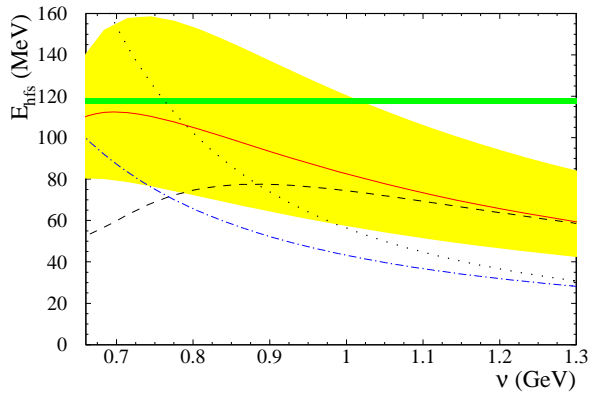,height=6cm}
\caption{\label{fig2} HFS of 1S charmonium as a function of the
renormalization scale $\mu$ in the LO (dotted line), NLO (dashed line), LL
(dot-dashed line), and NLL (solid line) approximations. 
For the NLL result, the band reflects the errors
due to $\alpha_s(M_Z)=0.118\pm 0.003$. The horizontal band gives 
the experimental value $117.7\pm 1.3$~MeV \cite{Hag}.}
\end{figure}

Another way to estimate the accuracy of our result is by considering
the HFS in charmonium, where experimental data are
available. The result of our analysis is
given in  Fig.~\ref{fig2} along with the experimental value 
$117.7\pm 1.3$~MeV \cite{Hag}. The local maximum
of the NLL curve corresponds to $E_{\rm hfs}=112$~MeV.
Taking into account that, in charmonium, the convergence of the perturbative
and relativistic expansion is much worse and the nonperturbative
effects are much less suppressed in comparison to the bottom-quark case, the
good agreement between our perturbative result 
and the experiment is
amazing and supports our error estimates.
We should emphasize
the crucial role of the resummation to bring the perturbative 
analysis in agreement with the experimental data.  
Note that the recent lattice estimates undershoot the experimental
value by $20-30\%$ \cite{Choe}.

We may also apply our formulae to the $n=2$ excited states. For bottomonium,
a recent study of unquenched nonrelativistic lattice QCD with $1+2$
light flavors by the HPQCD and UKQCD collaborations \cite{Gra} predicts
the ratio $E_{\rm hfs}(2S)/E_{\rm hfs}(1S)=0.32\pm 0.07$, which is in good
agreement with the value 0.27 obtained using our NLL result. 
For charmonium, our perturbative estimate
$E_{\rm hfs}(2S)/E_{\rm hfs}(1S)=0.41$  also  agrees wery well
with the result $0.41\pm 0.03$ of the recent experimental measurements 
\cite{Cho}. Although one definitely cannot rely on the (even NRG-improved) 
perturbative analysis  of the excited charmonium
states, the above agreement suggests that the
nonperturbative effects are not dominant and well under control
at least for the ground state. 

In principle, our result can also be applied to the analysis of the
difference between the resonance energies in $e^+e^-\to t\bar t$ and
$\gamma\gamma\to t\bar t$ threshold production \cite{PenPiv}, which is
beyond the scope of this Letter.

To conclude, we have computed the heavy-quarkonium HFS in the NLL
approximation by summing up the subleading logarithms 
of $\alpha_s$ to all orders in the perturbative expansion. The use
of the NRG extends the range of $\mu$ where the perturbative result
is stable to the physical scale of the inverse Bohr radius. 
We found the resummation of logarithms to be crucial to reach
agreement between our perturbative estimate and the experimental data on
the HFS in charmonium despite {\it a priori} unsuppressed nonperturbative
effects. Our results further indicate  that the properties of 
the physical charmonium and bottomonium  ground states
are dictated by perturbative  dynamics. 
As an application of the result to the bottomonium spectrum,
we predict the mass of the as-yet undiscovered $\eta_b$ meson to be
\begin{equation}
M(\eta_b)=9419\pm 11\,{(\rm th)} \,{}^{+9}_{-8}\, 
(\delta\alpha_s)~{\rm MeV}\,,
\end{equation}
where the errors due to 
the high-order perturbative corrections and the nonperturbative effects
are added up in quadrature in ``th'',  whereas ``$\delta\alpha_s$''
stands for the uncertainty in $\alpha_s(M_Z)=0.118\pm0.003$.
If the experimental error in
future measurements of  $M(\eta_b)$ will not
exceed  a few MeV, the bottomonium HFS will become a  competitive source 
of  $\alpha_s(M_Z)$ with an estimated accuracy
of  $\pm 0.003$, as can be seen from Fig.~\ref{fig1}.

\begin{acknowledgments}
This work  was supported in part by DFG Grant
No.\ KN~365/1-1 and BMBF Grant No.\ 05~HT1GUA/4. 
The work of A.A.P. was supported in part by BMBF Grant No.\ 05HT4VKA/3
and SFB Grant No. TR 9. The work of A.P. was
supported in part by by MCyT and Feder (Spain), FPA2001-3598, by CIRIT
(Catalonia), 2001SGR-00065 and by the EU network EURIDICE,
HPRN-CT2002-00311. The work of V.A.S. was supported in part by 
RFBR Project No. 03-02-17177, 
Volkswagen Foundation Contract No. I/77788, and DFG Mercator 
Visiting Professorship No. Ha 202/110-1.

\end{acknowledgments}


\begin{thebibliography}{99}

\bibitem{AppPol} T. Appelquist and H.D. Politzer,
Phys.\ Rev.\ Lett.\ {\bf34}, 43 (1975).

\bibitem{KPP} J.H. K\"uhn, A.A. Penin, and A.A. Pivovarov,
Nucl.\ Phys.\ {\bf B534}, 356 (1998);
A.A. Penin and A.A. Pivovarov,
Phys.\ Lett.\ B {\bf435}, 413 (1998);
Nucl.\ Phys.\ {\bf B549}, 217 (1999);
K. Melnikov and A. Yelkhovsky,
Phys.\ Rev.\ D {\bf59}, 114009 (1999);
M. Beneke and  A. Signer,  
Phys.\ Lett.\ B {\bf471}, 233 (1999); 
J.H. K\"uhn and  M. Steinhauser,  
Nucl.\ Phys.\  {\bf B619}, 588 (2001); 
{\bf B640},  415(E) (2002);
A.H. Hoang, Report No. CERN-TH-2000-227 and
hep-ph/0008102.

\bibitem{PinYnd}  A. Pineda and F.J. Yndurain,
Phys.\ Rev.\ D {\bf 58},  094022 (1998);
A. Pineda, JHEP\ {\bf0106}, 022 (2001);
N. Brambilla, Y. Sumino, and A. Vairo, 
Phys.\ Rev.\ D {\bf 65}, 034001 (2002). 

\bibitem{PenSte} A.A. Penin and M. Steinhauser,
Phys.\ Lett.\ B {\bf538}, 335 (2002).

\bibitem{Hei} ALEPH Collaboration, A. Heister {\it et al.}, 
Phys.\ Lett.\ B {\bf530}, 56 (2002). 

\bibitem{Sto} H. St\"ock, hep-ex/0310021.

\bibitem{BucNg} 
W. Buchm\"uller, Y.J. Ng, and S.H.H. Tye,  
Phys.\ Rev.\ D {\bf24}, 3003 (1981);
S.N. Gupta, S.F. Radford, and W.W. Repko 
Phys.\ Rev.\ D {\bf26}, 3305 (1982);
J. Pantaleone, S.H.H. Tye, and Y.J. Ng,  
Phys.\ Rev.\ D {\bf33}, 777 (1986);
S. Titard and F.J. Yndurain,
Phys.\ Rev.\ D {\bf49}, 6007 (1994).

\bibitem{CasLep} W.E. Caswell and G.P. Lepage,
Phys.\ Lett.\ B {\bf167}, 437 (1986);
G.T. Bodwin, E. Braaten, and G.P. Lepage,
Phys.\ Rev.\ D {\bf51}, 1125 (1995); {\bf55}, 5853(E) (1997).

\bibitem{PinSot1} A. Pineda and J. Soto,
Nucl.\ Phys.\ B (Proc.\ Suppl.) {\bf64}, 428 (1998).

\bibitem{Pin} A. Pineda,
Phys.\ Rev.\ D {\bf65}, 074007 (2002); 
{\bf66}, 054022 (2002).

\bibitem{HMS} A.H. Hoang, A.V. Manohar, and I.W. Stewart,
Phys.\ Rev.\ D {\bf 64}, 014033 (2001); 
A.H. Hoang and I.W. Stewart,
Phys.\ Rev.\ D {\bf 67}, 014020 (2003).

\bibitem{LMR} M.E. Luke, A.V. Manohar, and I.Z. Rothstein,
Phys.\ Rev.\ D {\bf61}, 074025 (2000).

\bibitem{BenSmi} M. Beneke and V.A. Smirnov,
Nucl.\ Phys.\ {\bf B522}, 321 (1998);
V.A. Smirnov,
{\it Applied Asymptotic Expansions in Momenta and Masses}
(Springer-Verlag, Heidelberg, 2001).

\bibitem{KPSS} B.A. Kniehl, A.A. Penin, 
V.A. Smirnov, and M. Steinhauser, 
Phys.\ Rev.\ D {\bf65}, 091503(R) (2002);
Nucl.\ Phys.\ {\bf B635}, 357 (2002);
Phys.\ Rev.\ Lett.\ {\bf90}, 212001 (2003); 
{\bf91}, 139903(E) (2003).

\bibitem{PinSot2} A. Pineda and J. Soto,
Phys.\ Lett.\ B {\bf420}, 391 (1998);
Phys.\ Rev.\ D {\bf59}, 016005 (1999).

\bibitem{CMY} A. Czarnecki, K. Melnikov, and A. Yelkhovsky,
Phys.\ Rev.\ A {\bf59}, 4316 (1999).

\bibitem{KniPen1} B.A. Kniehl and A.A. Penin,
Nucl.\ Phys.\ {\bf B563}, 200 (1999);
N. Brambilla, A. Pineda, J. Soto, and A. Vairo,
Phys.\ Lett.\ B {\bf 470}, 215 (1999).

\bibitem{ABN} G. Amor\'os, M. Beneke, and M. Neubert, 
Phys.\ Lett.\ B {\bf 401}, 81 (1997).

\bibitem{KniPen2} B.A. Kniehl and A.A. Penin,
Nucl.\ Phys.\ {\bf B577}, 197 (2000).


\bibitem{Manohar} A.V. Manohar,  
Phys.\ Rev.\ D {\bf 56}, 230 (1997).
 
\bibitem{VolLeu} M.B. Voloshin,
Nucl.\ Phys.\ {\bf B154}, 365 (1979);
Yad.\ Fiz.\ {\bf36}, 247 (1982)
[Sov.\ J. Nucl.\ Phys.\ {\bf36}, 143 (1982)];
H. Leutwyler,
Phys.\ Lett.\ B {\bf98}, 447 (1981).


\bibitem{TitYnd2} S. Titard and F.J. Yndurain,
Phys.\ Rev.\ D {\bf51}, 6348 (1995);
A. Pineda,
Nucl.\ Phys.\ {\bf B494}, 213 (1997).

\bibitem{Eic} SESAM Collaboration, N. Eicker {\it et al.}, 
Phys.\ Rev.\ D {\bf57}, 4080 (1998).

\bibitem{Man} CP-PACS Collaboration, T. Manke {\it et al.}, 
Phys.\ Rev.\ D {\bf62}, 114508 (2000).  
We quote the result of the lattice simulation which fits the
experimental values of the $\Upsilon(1S)$ mass and the ratio of
the $2^3S_1-1^3S_1$ and $^1P_1-1^3S_1$ splittings.

\bibitem{Lia} X. Liao and T. Manke, Phys.\ Rev.\ D {\bf65}, 074508
(2002).  We quote the result of the lattice simulation which fits the
experimental value of the $\Upsilon(1S)$ mass.

\bibitem{Hag} K. Hagiwara {\it et al.},
Phys.\ Rev.\ D {\bf 66}, 010001 (2002).

\bibitem{Choe} QCD-TARO Collaboration, S. Choe {\it et al.}, JHEP {\bf0308}, 022 (2003);
MILC Collaboration, M. di Pierro {\it et al.},  hep-lat/0310042.

\bibitem{Gra} HPQCD Collaboration, A. Gray {\it et al.},  Nucl.\ Phys.\
B (Proc.Suppl.) {\bf 119}, 592 (2003).
The quoted value is extracted from Fig.~3.

\bibitem{Cho} We take the average of 
the results reported in BELLE Collaboration, S.-K. Choi {\it et al.},
Phys. Rev. Lett. {\bf 89}, 102001 (2002); {\bf 89}, 129901(E) (2002);
BELLE Collaboration, K. Abe {\it et al.}, 
Phys. Rev. Lett. {\bf 89}, 142001 (2002);
CLEO Collaboration, J. Ernst {\it et al.},
Report No.~CLEO-CONF-03-05 and hep-ex/0306060; 
BABAR Collaboration, 
B. Aubert {\it et al.}, Report No.~BABAR-PUB-03/023 and  
hep-ex/0311038.

\bibitem{PenPiv} A.A. Penin and A.A. Pivovarov,
Nucl.\ Phys.\ {\bf B550}, 375 (1999);
Yad.\ Fiz.\ {\bf64}, 323 (2001)
[Phys.\ Atom.\ Nucl.\ {\bf64}, 275 (2001)].

\end{thebibliography}
\end{document}